\documentclass[twoside]{LCWS11}
\usepackage[latin1]{inputenc}
\usepackage{graphicx}
\usepackage{wrapfig,rotating}
\usepackage{amssymb,amsmath,array}
\usepackage{cite}

\begin{document}
\title{
Top Higgs Yukawa Coupling Analysis from \\ $e^+e^- \rightarrow \bar{t} tH \rightarrow \bar{b} W^- bW^+\bar{b}b$} 
\author{Hajrah Tabassam$^1$ Victoria Martin$^2$
\vspace{.3cm}\\
1- Department of Physics\\
Quaid-i-Azam University Islamabad - Pakistan\\
\vspace{.1cm}\\
2-School of Physics and Astronomy  \\
University of Edinburgh  - United Kigdom
}

\maketitle
\begin{abstract}
Electroweak symmetry breaking and the origin of mass of fermion and boson are fundamental questions to understand particle physics. The essential piece of this symmetry breaking, the Higgs boson, is the most probable particle to be discovered at the Large Hadron Collider (LHC).  
If one, or more, Higgs particles are discovered, 
precise measurements of all the properties of this Higgs will be very important including the measurement of Yukawa couplings of the fermions to the Higgs.
We present a study of $e^+e^- \rightarrow t\bar{t}H$ at a linear collider, with the aim of making a direct measurement of the the top-Higgs coupling, $g_{t\bar t H}$ using the semi-leptonic final state and $M_H$ of 120 GeV. We show that the top-Higgs coupling at the center of mass energy 500 GeV, can be measured with an accuracy of better than 28\%.
\end{abstract}

\section{Introduction}
The ILC is a unique tool which could play an extremely important role in high-precision measurements. Data from the ILC will allow a determination of the profile of the Higgs bosons and their fundamental properties with a high level of confidence and will provide a unique opportunity to establish experimentally the mechanism that generates the particle masses.

The Higgs couplings to the fermions and the gauge bosons are related to the masses of these particles and the only free parameter of the model is the mass of the Higgs boson itself; there are, however, both experimental and theoretical constraints on the Higgs mass in the Standard Model.
Experimentally, the available direct information on the Higgs mass ($M_H$) is the lower limit $M_H \gtrsim 114.4$ GeV established at LEP \cite{Barate:2003sz}.  
The high accuracy of the electroweak data measured at LEP, SLC and Tevatron provides an indirect sensitivity to $M_H$: the Higgs boson contributes logarithmically, $\propto \log(M_H/M_W)$, to the radiative corrections to the $W/Z$ boson propagators. A recent analysis, which uses the updated value of the top quark mass yields the value 85$\pm$25 GeV, with upper limit of $M_H\lesssim165$ GeV~\cite{Erler:2010wa}. 
 Direct searches by CDF and D0 exclude a region at high mass between $156 < M_H < 177$ GeV~\cite{ref2.6}.   
ATLAS and CMS have also made direct searches for the Higgs boson.  At the time of this workshop preliminary LHC results exclude the ranges 155-190 GeV~\cite{ATLAS} at ATLAS and 149-206 GeV~\cite{CMS} at CMS at 95\% CL. 

\section{Higgs Production and Decay}
In the Standard Model (SM), the profile of the Higgs particle is uniquely determined once its mass $M_H$ is fixed~\cite{ref2,ref3}. The decay width, the branching ratios and the production cross sections are given by the strength of the Yukawa couplings to fermions and gauge bosons, the scale of which is set by the masses of these particles.
\begin{displaymath}
g_{fHf} \propto m_f
\end{displaymath}
For a top quark mass of $M_t=175\;\mathrm{GeV}$, the Standard Model the Higgs Yukawa coupling is ${g_{t\bar{t}H}}=M_t/v=0.71$, where $v=246\;\mathrm{GeV}$ is the vacuum expectation value.
The aim of the work presented here is to determine how accurately ${g_{t\bar{t}H}}$ can be determined at the ILC through a measurement of $t \bar t H$ production.

\begin{wrapfigure}{r}{0.5\columnwidth}
\centerline{\includegraphics[width=0.45\columnwidth]{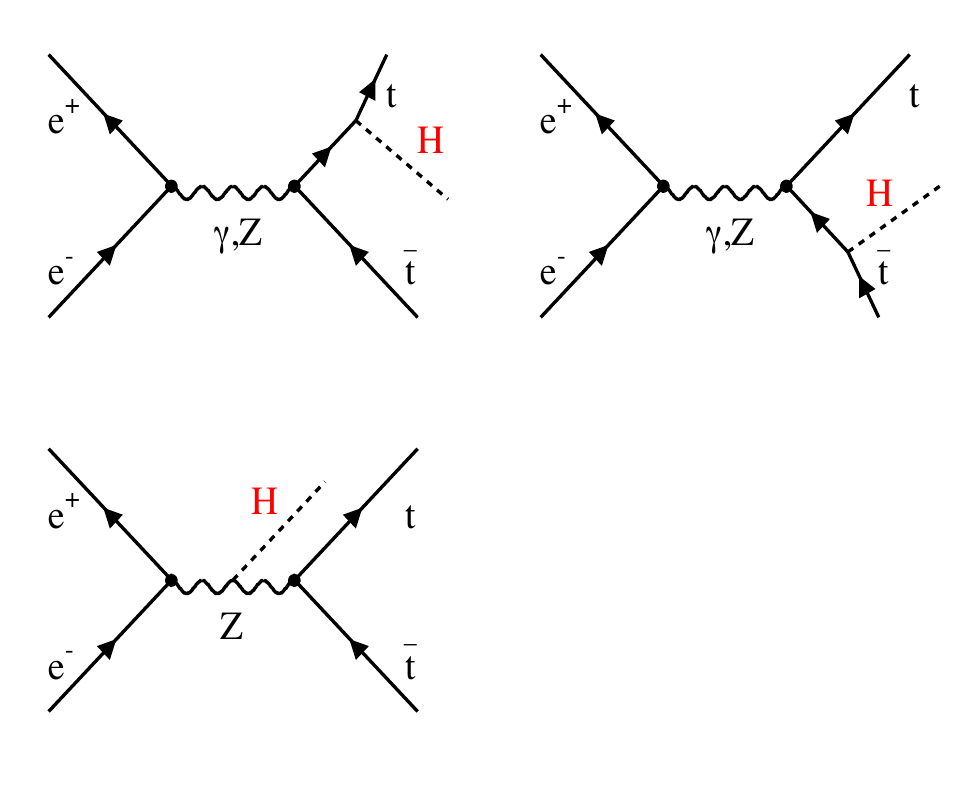}}
\caption {Feynman diagrams for Higgs production in association with $t \bar{t}$ pair.}\label{topHFD}
\end{wrapfigure}

There are two Higgs main production channels for $e^+e^-$ production at $\sqrt{s}=500\;\mathrm{GeV}$: the Higgs-strahlung process including $e^+e^- \rightarrow ZH$ and associated production with top pair $e^+e^-\rightarrow t \bar t H$ (figure~\ref{topHFD}) which we use for our analysis. The diagram where the Higgs boson is radiated from the $Z$ boson modifies only slightly the cross-section and thus are neglected in this study.\\

The signal has a cross section of $0.576\;\mathrm{fb}$ at $\sqrt {s} = 500$ GeV due to phase space suppression but, at $\sqrt{s} = 800$ GeV, it can reach the level of a few femtobarns.

\section{Overview}
References~\cite{ref6.2} and~\cite{ref6.3} have performed earlier feasibility studies of the top-Higgs Yukawa coupling via the process $e^+ e^- \rightarrow t\bar{t}H$ at a linear collider and in~\cite{ref6.1}, an analysis was performed at 800 GeV center-of-mass energy.  
For the ILD detector, we perform the first complete analysis of the process for $M_H=120\;\mathrm{GeV}$ in the semileptonic channel where one $W$ boson from the top quark decay, $t\rightarrow bW$, decays as $W \rightarrow \ell \nu $ and the other as $W \rightarrow q\bar{q}$. A full reconstruction of the final state is conducted by reconstructing one hadronic and one leptonic $W$ decays. The top quarks and the Higgs boson are then reconstructed from $W'$s and $b$-jets. Missing energy reconstruction has a large impact on the reconstruction of the semi-leptonic channel, and $b$-tagging has a crucial role in separating signal and background. The analysis is carried out for an integrated luminosity, $\mathcal L=1000\;\mathrm{fb}^{-1}$ and at $\sqrt{s} = 500$ GeV. The main background processes are $t\bar{t}$ and $t\bar{t}Z$ production. 

\section{Monte Carlo Samples} 
The analysis was performed with the samples provided by the ILD optimisation group~\cite{ref2.7} with $M_H=120\;\mathrm{GeV}$ and $M_t=175\;\mathrm{GeV}$. 
The samples were generated at SLAC using WHIZARD, and then simulated and reconstructed on the DESY Grid nodes. The GEANT4 based Mokka package was used for simulation in the ILD-00 detector model, which is the first simulated reference model of ILD. Reconstruction was performed using MarlinReco, PandoraPFA and LCFIVertex with the versions supplied in ILCInstall \texttt{v01-07}.
 
 In table~\ref{Sgbkg6}, the sample sizes used in the analysis are shown. \\
\begin{table}[htdp]
\caption{Cross section and luminosity for signal and background processes.}
\label{Sgbkg6}
\begin{center}
\begin{tabular}{|c|c|r|c|}
\hline
Process & $\sigma$ (fb) & events & $\mathcal L$ (ab$^{-1}$)\\ \hline
$e^+  e^- \rightarrow t\bar{t} H$ & 0.576 & 20000 & 34 \\ \hline
$e^+  e^- \rightarrow t\bar{t}\rightarrow \ell\nu q\bar{q}$ & 230 & 400000 & 1.7 \\ \hline
$e^+  e^- \rightarrow t\bar{t} Z$ & 0.58 & 24000 & 41\\ \hline

\end{tabular}
\end{center}
\label{default}
\end{table}%

\section{Event Reconstruction}
We reconstruct the semi-leptonic final state of the signal process with one charged lepton (electron or muon) and a neutrino and six jets, accounting for 
$\sim20\%$ of the total cross section.
The final state in this channel has four $b$-jets, two light jets, $j$, one charged lepton, $\ell$, and missing energy for $\nu$:
\begin{displaymath}
e^+e^- \rightarrow t\bar{t}H \rightarrow W^+bW^-\bar{b} b\bar{b} \rightarrow \ell\nu\,2j\,4b
\end{displaymath}

\begin{figure}[tbp]
\begin{center}
\includegraphics[width=6.25cm]{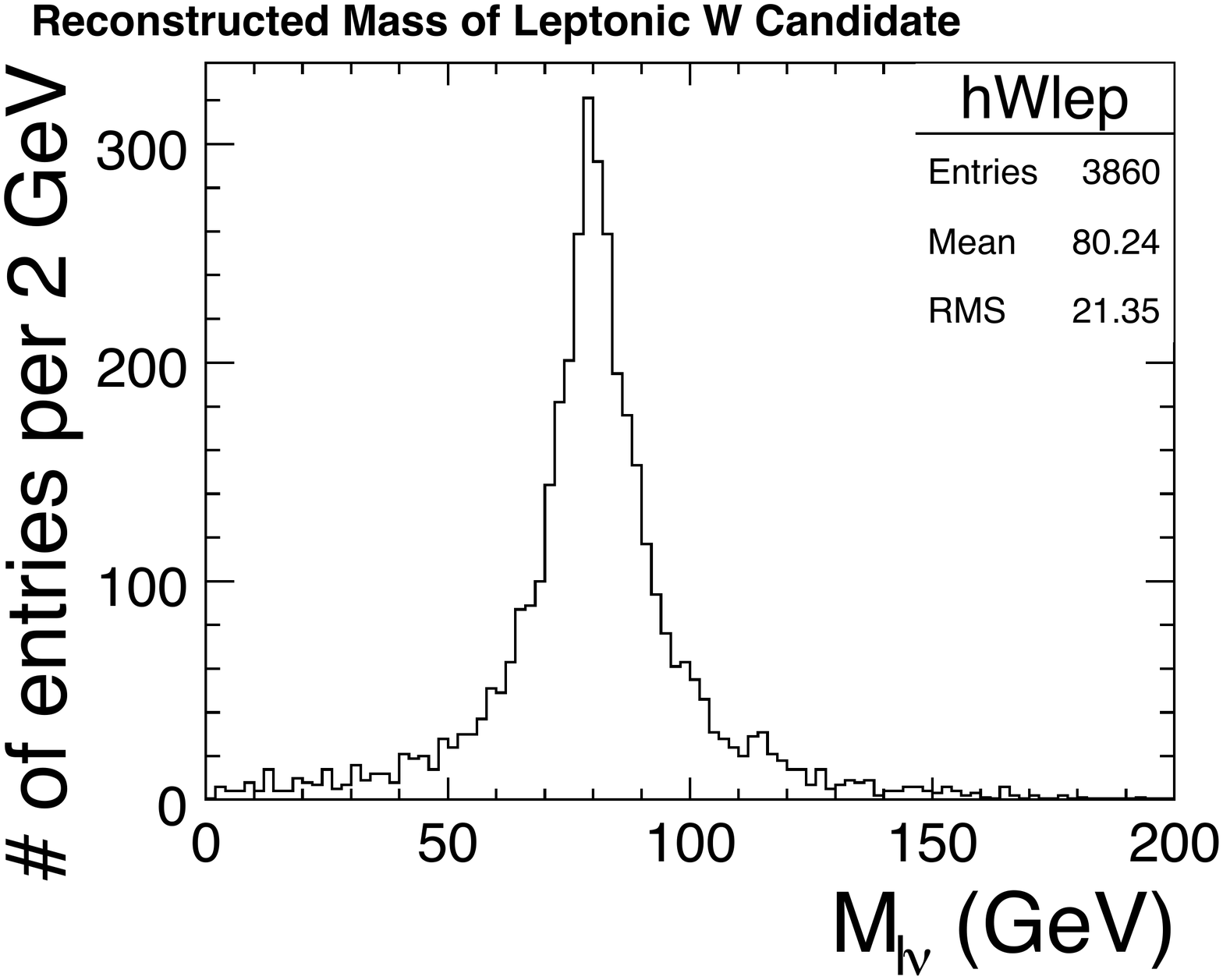}
\includegraphics[width=6.25cm]{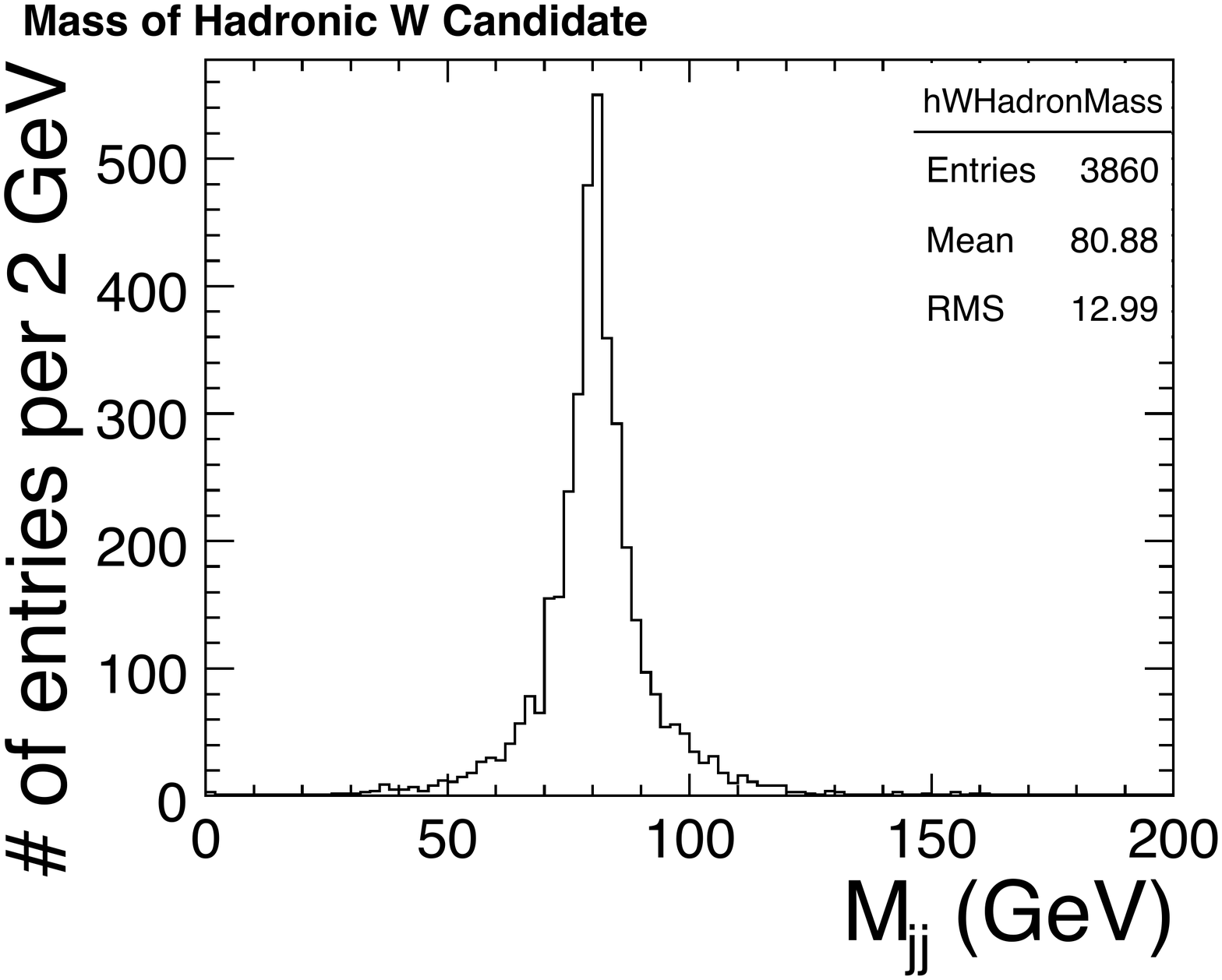}
\caption{\label{7.7}{Left: Reconstructed mass of leptonic $W$ candidate combining reconstructed lepton momentum and missing energy. Right: mass of hadronic $W$ candidate reconstructed by selecting the light di-jets pair with an invariant mass closest to $M_W$.}}
\end{center}
\end{figure}

We select electrons and muons using selection criteria obtained from a single particle sample study.
Left figure~\ref{7.7} shows the reconstructed invariant mass of the charged lepton and the missing energy in the event. 

We remove the electrons and muons from our sample and force the remaining particles into exactly six jets using the JETFinder algorithm. The LCFIVertex algorithm is run on the jets to obtain $b$-tagging information~\cite{ref4.8}. 

Jets passing the LCFIVertex reconstruction are sorted according to their $b$-tag value.  The $b$-tag uses a neural net algorithm, producing a value between 0 and 1 for each jet: 1 being the most $b$-jet like, and 0 being the least $b$-jet like.  To suppress backgrounds, and improve the reconstruction efficiency, we select events where at least four of the six jets have $b$-tag $>0.09$. 
The four jets with the highest $b$-tag value are considered as the $b$-jets and the two with the lowest $b$-tag value are considered as the light jets. Hadronic $W$ candidates, shown in right figure~\ref{7.7}, are created from pairs of light-jets in the event. 

The remaining jets are considered as the $b$-jets and used to reconstruct $M_{bb}, M_{\ell\nu b}$ and $M_{jjb}$. 

Top-quark candidates are reconstructed using a tagged $b$-jet and a reconstructed $W$ candidate and the Higgs boson is reconstructed from a pair of tagged $b$-jets. All possible combinations of tagged $b$-jets are examined to reconstruct the Higgs boson and both top-quarks simultaneously. 
\begin{figure}[tbp]
\begin{center}
\includegraphics[width=6.25cm]{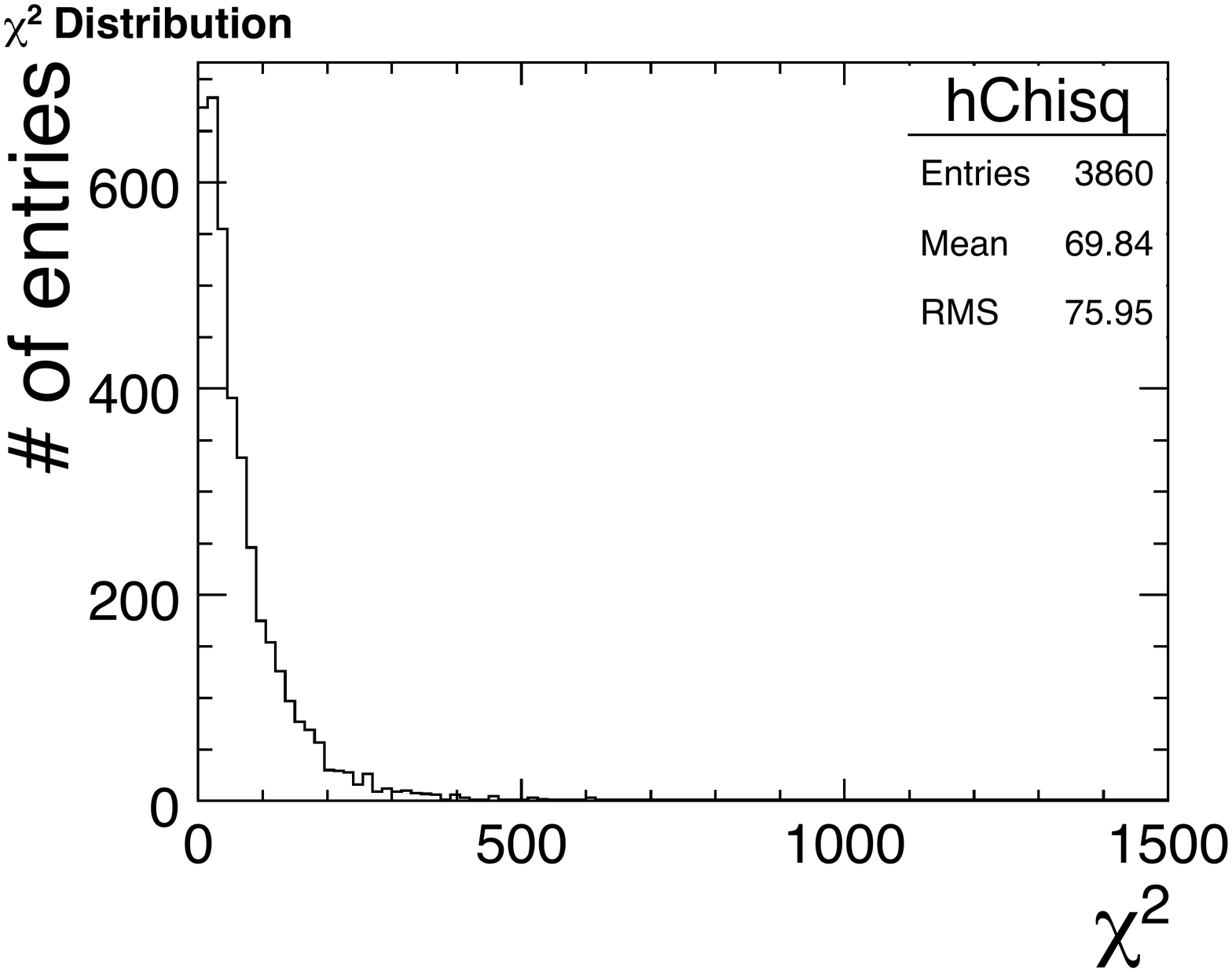}
\includegraphics[width=6.25cm]{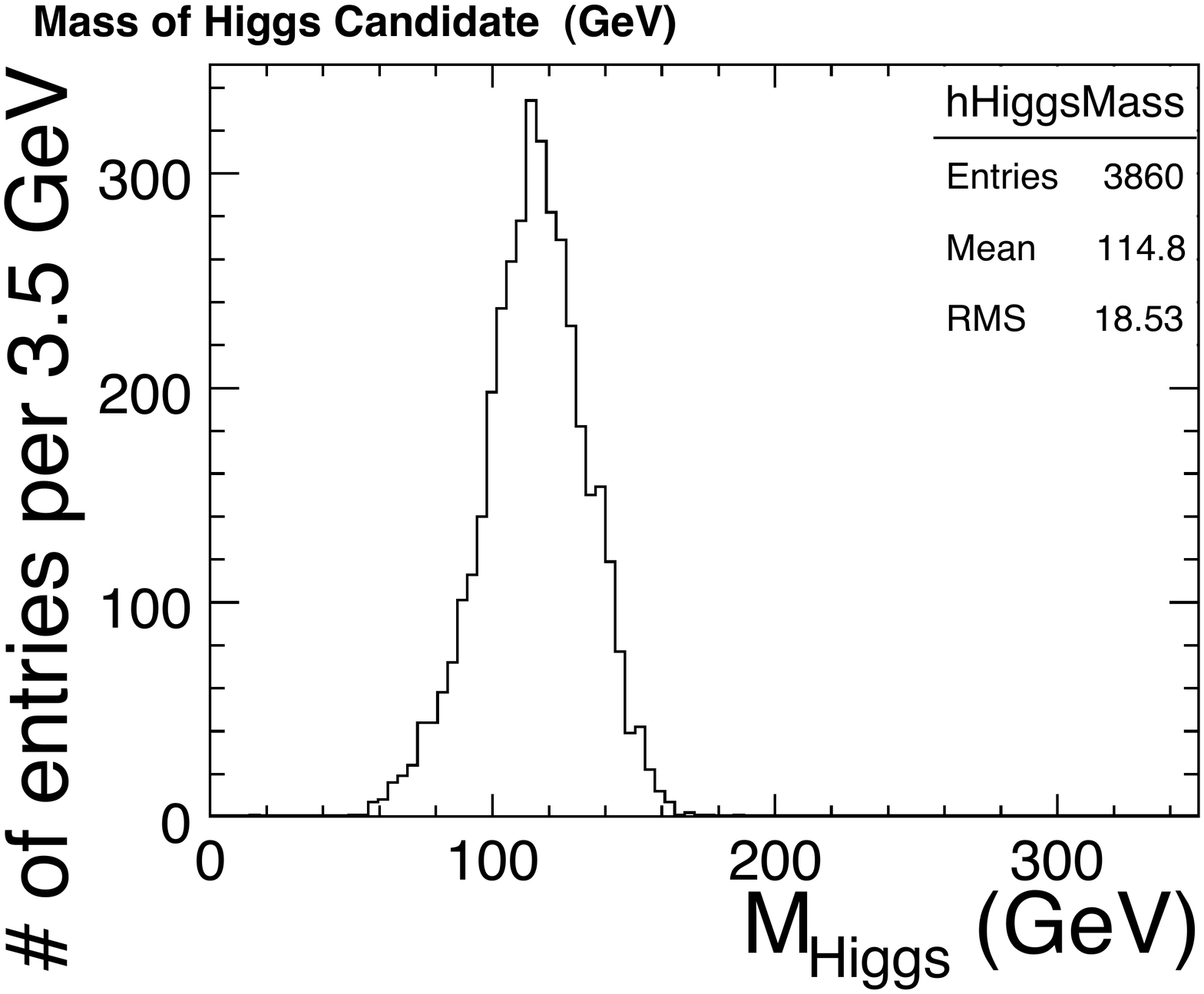}
\includegraphics[width=6.25cm]{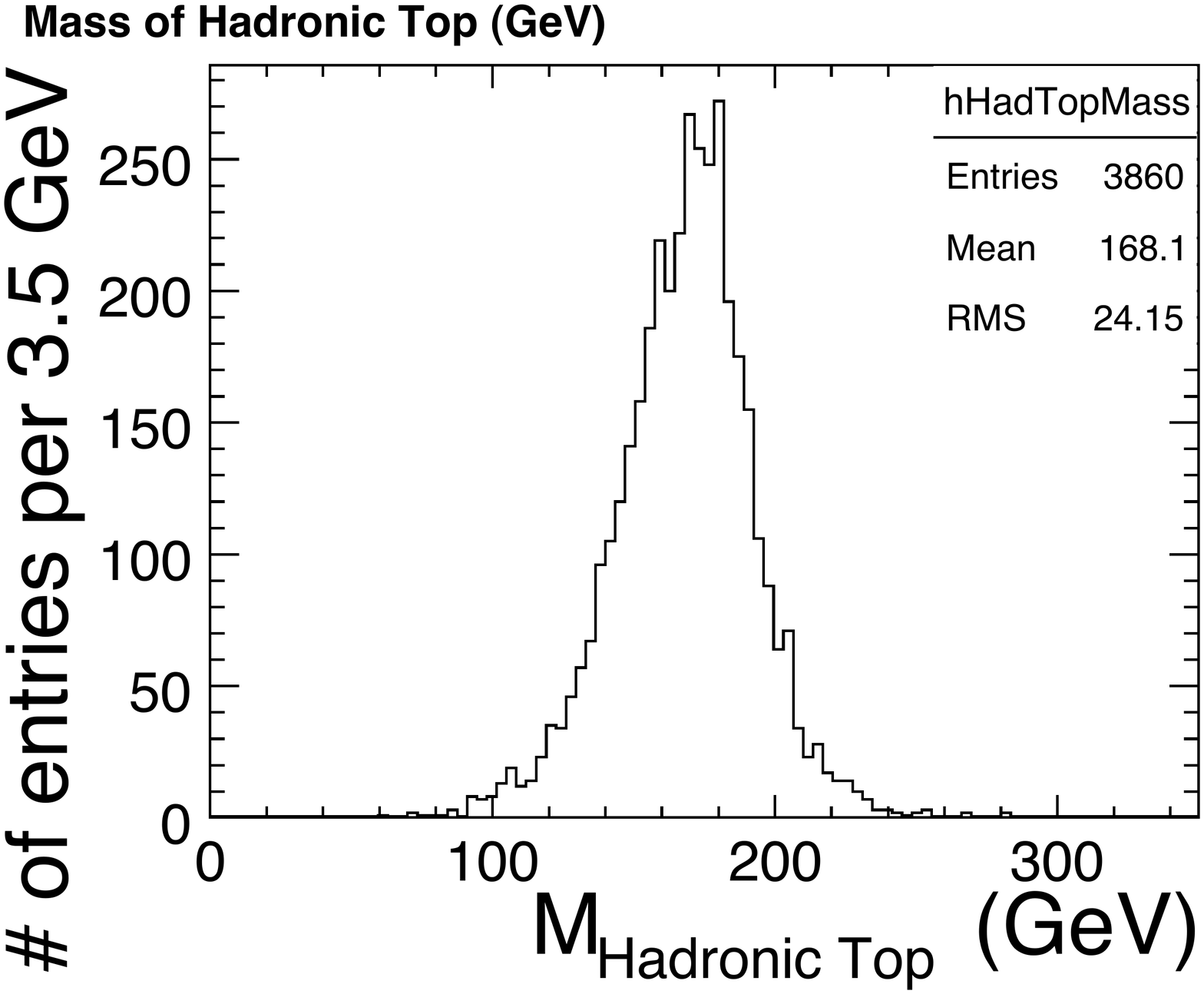}
\includegraphics[width=6.25cm]{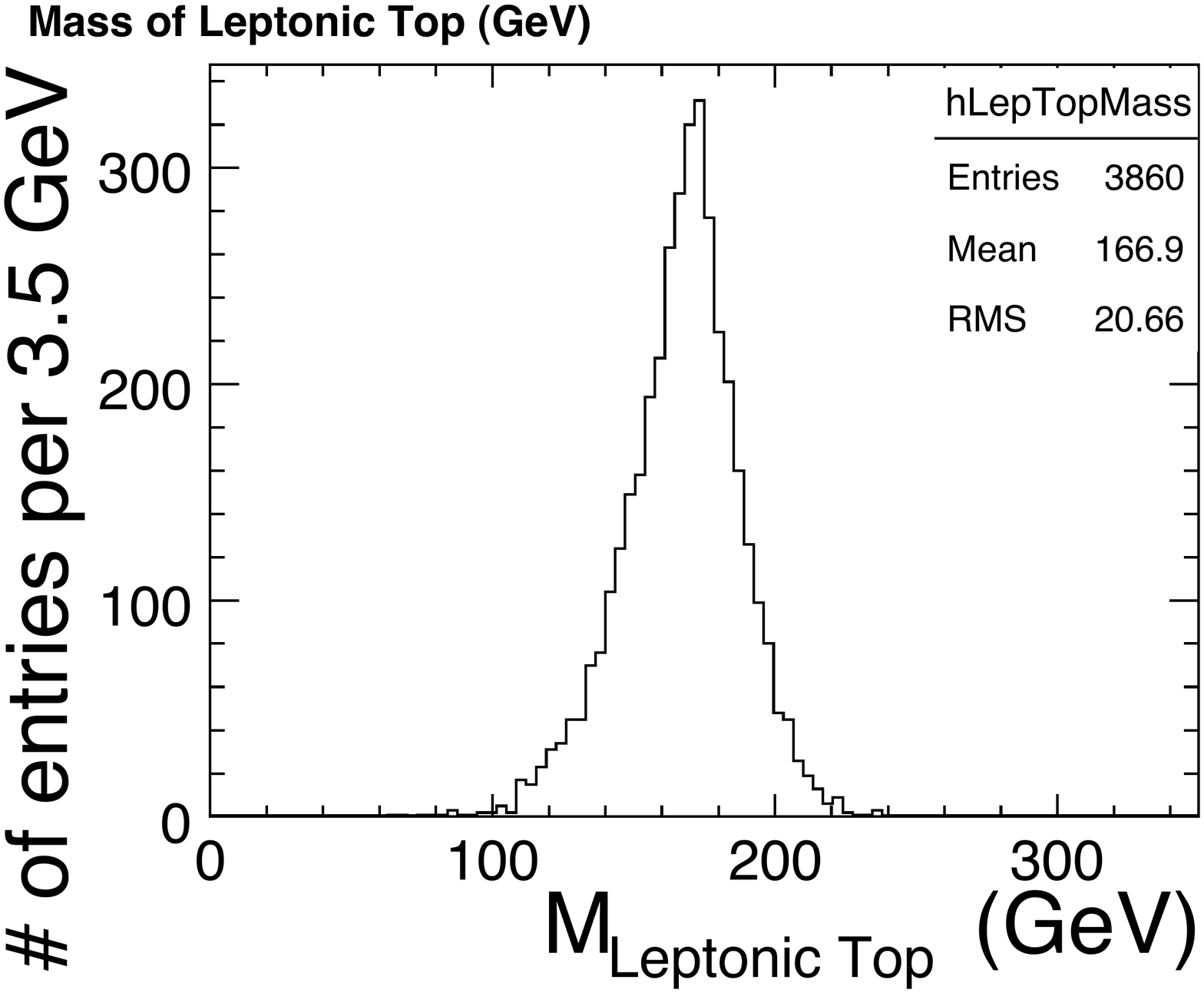}
\caption[Plots for the best Higgs and top quarks combination.]{\label{7.10}{Plots for the best Higgs and top quarks combination. Top left: $\chi^2$,  
Top right: Reconstructed mass of b-jet pair, $M_{bb}$. Bottom left: Reconstructed mass of two light- and one b-jet, $M_{jjb}$. Bottom right: Reconstructed mass of lepton, missing energy and b-jet, $M_{\ell\nu b}$.}}
\end{center}
\end{figure}

To reduce combinatorial background, we define $\chi^2$ for each event:
\begin{equation}
\label{eq:chi}\chi ^2 = \frac{(M_{\ell\nu b} - M_t)^2}{\sigma_{\ell\nu b} ^2} + \frac{(M_{jj b} - M_t)^2}{\sigma_{jjb} ^2} + \frac{(M_{b b} - M_{H})^2}{\sigma_{b b} ^2}\end{equation}
where take the widths, $\sigma$, for each combinations from the reconstructed resolutions:

\begin{displaymath}
\sigma_{\ell\nu b} = 20.7 \pm 0.2 \;\mathrm{GeV}\hspace{0.5cm}
\sigma_{jjb} = 23.5 \pm 0.3 \;\mathrm{GeV}\hspace{0.5cm}
\sigma_{b b} = 19.3 \pm 0.2\;\mathrm{GeV}
\end{displaymath}

The mass distributions of the lowest $\chi^2$ combination are given in figure~\ref{7.10}.

\section{Signal and Background Separation}
\begin{figure}[tbp]
\begin{center}
\includegraphics[width=14cm] {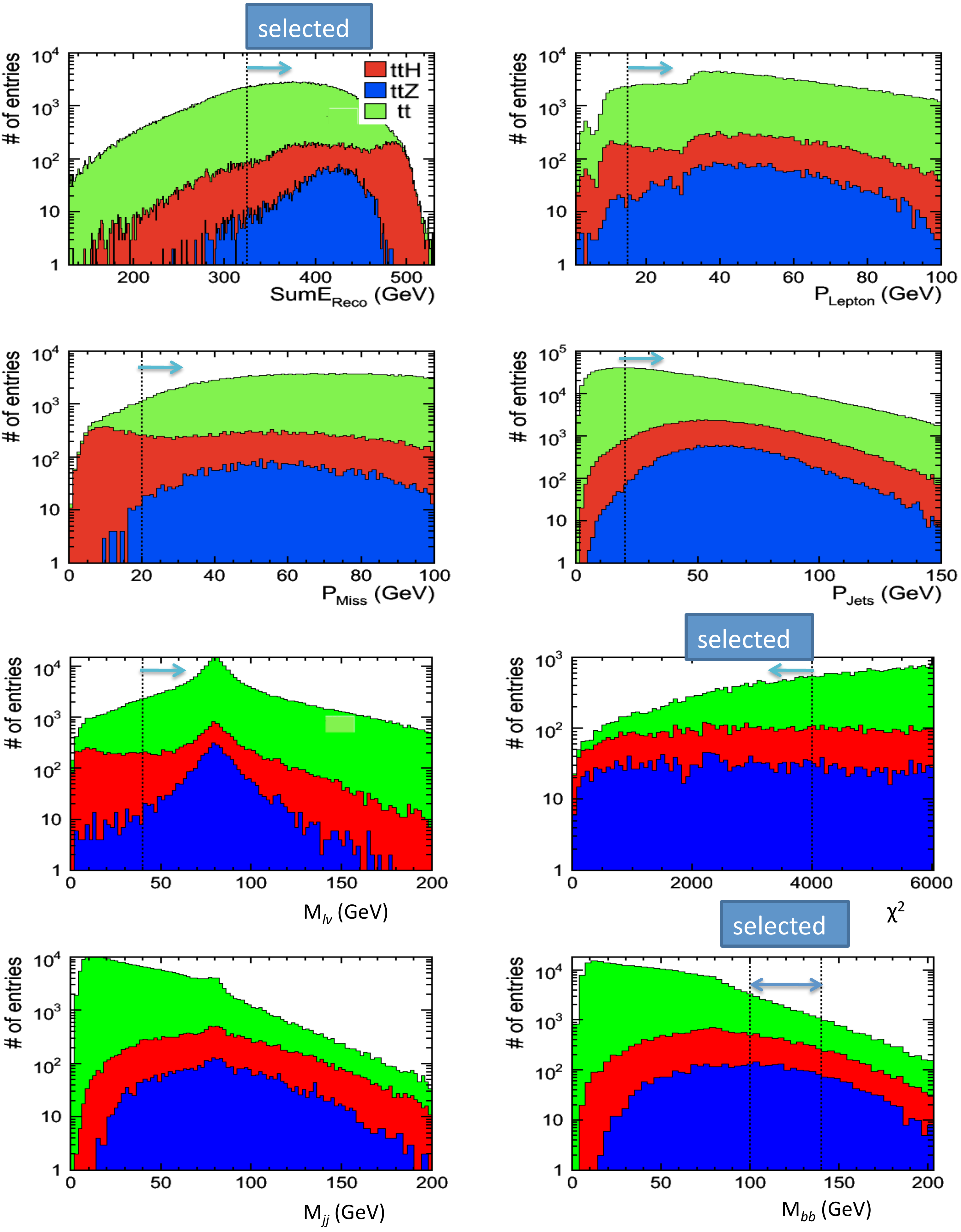}
\caption{\label{7.12}{Stacked plots showing selection variables for signal and background. The dotted lines show the cut values selected.}}
\end{center}
\end{figure}

The main backgrounds are from $t\bar{t}$ and $t\bar{t}Z$.  We used the following variables to discriminate between signal and background by maximising ${S}/{\sqrt{S+B}}$ (where $S$ is signal and $B$ is background).
\begin{itemize}
\item$E_{\mathrm{SumReco}}$: sum of the energies of all reconstructed particles in the event. 
\item$P_{{\mathrm{Lepton}}}$: the reconstructed momentum of the identified lepton.
\item$P_{{\mathrm{miss}}}$: the reconstructed missing momentum.  
\item$P_{{\mathrm{Jets}}}$: the momentum of each of the reconstructed jets.
\item $\chi^2$, as defined in equation~(\ref{eq:chi}).
\item $m_{\ell\nu}$: leptonic $W$ candidate mass 
\item $m_{b\bar b}$: Higgs candidate mass.
\item Total reconstructed mass $m_{b\bar b}+m_{\ell\nu b}+m_{jjb}$.
\end{itemize}
Figure~\ref{7.12} shows some of these variables for signal and background.

The optimised selection on the above variables selects  $(0.11\pm0.01)\%$ of $t\bar{t}$,  $(2.76\pm0.12)\%$ of $t\bar{t}Z$, and  $(7.57\pm0.19)\%$ of signal events. 

\section{Measurement of $\bold{g_{t\bar{t}H}}$}
In the Standard Model the Higgs Yukawa coupling is written in terms of the top quark mass ${g_{t\bar{t}H}}=M_t/v=0.71$.
To measure the top Higgs Yukawa coupling, we followed procedure given in~\cite{ref6.1}. 
The statistical and systematic uncertainties on the measurement of $g_{t\bar{t}H}$ can be expressed as:
\begin{displaymath}
\left(\frac{\Delta g_{t\bar{t}H}}{g_{t\bar{t}H}}\right)_{\mathrm{stat}} \approx \frac{1}{S_{\mathrm{stat}}(g^2_{t\bar{t}H})\sqrt{\epsilon\rho\mathcal L}} \end{displaymath}
\begin{displaymath}
\left(\frac{\Delta g_{t\bar{t}H}}{g_{t\bar{t}H}}\right)_{\mathrm{syst}} \approx \frac{1}{S_{\mathrm{syst}}(g^2_{t\bar{t}H})} \frac{1-\rho}{\rho} \frac{\Delta\sigma_{\mathrm{eff}}^{\mathrm{BG}}}{\sigma_{\mathrm{eff}}^{\mathrm{BG}}}\\
\end{displaymath}
Where $\epsilon=(7.57\pm0.19)\%$ is the efficiency to select the data and $\rho=(12.5\pm0.3)\%$ is the purity of the final selected sample.  We use an integrated luminosity of $\mathcal L= 1000\;\mathrm{fb}^{-1}$, equivalent three to four year's running of the ILC.
The largest unknown comes from the knowledge of the background cross sections, 
$\Delta\sigma_{\mathrm{eff}}^{\mathrm{BG}}/{\sigma_{\mathrm{eff}}^{\mathrm{BG}}}$, dominated by current knowledge of $t\bar t$ production. Following reference~\cite{ref6.1}, we choose a relative uncertainty of 5\%;  
an ongoing study~\cite{ref6.4} will allow a better estimate of this uncertainty.

Ignoring the small contribution from Higgs radiation off the $Z$ (as shown in figure~\ref{topHFD})
we can write:
\begin{displaymath}
\frac{d\sigma_{t\bar{t}H}}{d(g^2_{t\bar{t}H})} 
\approx \frac{\sigma_{t\bar{t}H}}{g^2_{t\bar{t}H}} =0.087\;\mathrm{fb}
\end{displaymath}
with $\sigma_{t\bar{t}H}=0.044\;\mathrm{fb}$ being the effective cross section $7.57\%$ of the total cross section and ${g_{t\bar{t}H}}=0.71$.

Then the sensitivity factors $S_{\mathrm{stat}}$ and $S_{\mathrm{syst}}$ simply depend on cross section and on the squared coupling:
\begin{displaymath}
S_{\mathrm{stat}}(g^2_{t\bar{t}H}) = \frac{1}{\sqrt{\sigma_{t\bar{t}H}}}\left|\frac{d\sigma_{t\bar{t}H}}{d(g^2_{t\bar{t}H})}\right|= \frac{\sqrt{\sigma_{t\bar{t}H}}}{g^2_{t\bar{t}H}}=1.50 \; \mbox{fb}^{1/2}\hspace{1cm} 
\end{displaymath}
\begin{displaymath}
S_{\mathrm{syst}}(g^2_{t\bar{t}H}) = \frac{1}{\sigma_{t\bar{t}H}}  \left|\frac{d\sigma_{t\bar{t}H}}{d(g^2_{t\bar{t}H})}\right|=\frac{1}{g^2_{t\bar{t}H}}=1.98
\end{displaymath}
The expected statistical and systematic uncertainties are therefore: \begin{displaymath}\left(\frac{\Delta g_{t\bar{t}H}}{g_{t\bar{t}H}}\right)_{\mathrm{stat}}=21.6\% \hspace{1cm} \left(\frac{\Delta g_{t\bar{t}H}}{g_{t\bar{t}H}}\right)_{\mathrm{sys}}=17.6\%\end{displaymath} Combining the  statistical and systematic uncertainties in quadrature we obtain an estimate on the overall uncertainty on ${g_{t\bar{t}H}}$ of 27.9\%.

\section{Results and Discussion}
\begin{figure}[tbp]
\begin{center}
\includegraphics[width=13.46cm]{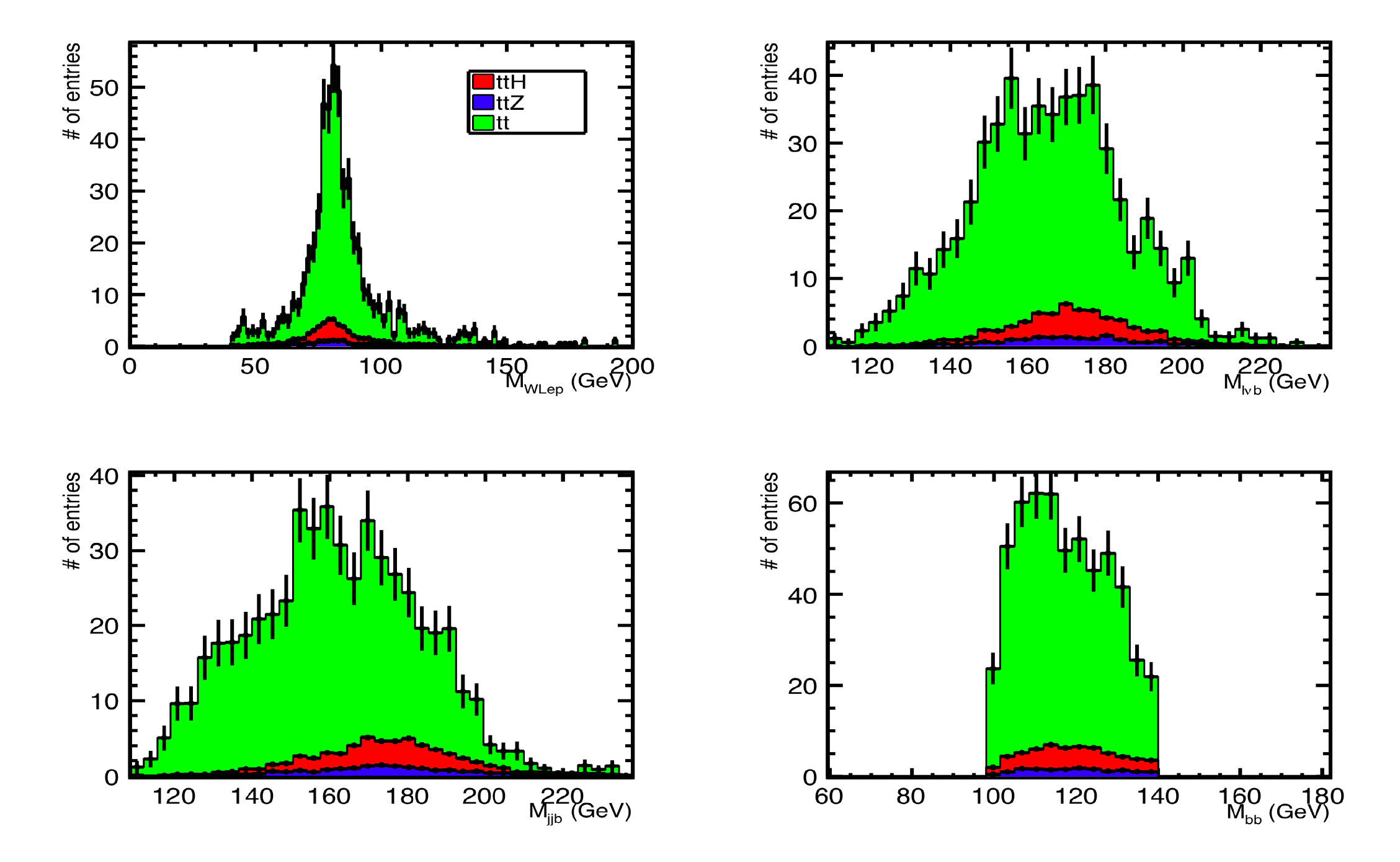}
\caption{\label{scaledplot}{After applying all selection cuts, the scaled signal and background distributions for Higgs and top masses. }}
\end{center}
\end{figure}

Figure~\ref{scaledplot} shows the distribution for the scaled signal and background samples. The main background after selection is due to the top-pair production.
The expected precision on measuring the top-Higgs Yukawa coupling is better than 28\% for $M_H = 120$ GeV, 
if the knowledge of $t\bar{t}$ background normalisation is known at the 5\% level. 
Much of the uncertainty on the results of this analysis is due to the small cross section of the $t\bar{t}H$ signal process. Even though the signal cross section is very small at the $\sqrt{s}$ = 500 GeV, yet 28\% accuracy is achieved. 
 Certainly there is room for improvement of the study presented.
  
In jet finding, we force particles into a jet which sometime don't belong to that jet. It could affect the energy of the reconstructed jet. A scaling of the flavour tag input variables to the reconstructed jet energy can minimise these effects. 

One of the areas of consideration is the improvement in the hadronic $W$ reconstruction.  
A cut on the $b$-tag value of the jets is used to identify light jets which are then combined to reconstruct the $W$ boson. This cut was decided to accommodate the selection cut on $b$-tag of third and fourth jet. A re-examining of the cut on $b$-tag of fifth and sixth jet could recover the signal events with a $b$-tag $>0.09$ and hence, improve the efficiency and purity of the sample. 

A cut based strategy is applied to discriminate the signal and background events. It is shown in previous studies that the use of neural network and likelihood methods can perform the signal background separation with better efficiency~\cite{ref6.3}. Hence, it is anticipated that a 2-3\% improvement  in the precision can be achieved by using neural network analysis. 

For the current study, six fermion backgrounds are not included but reference~\cite{ref6.1} shows that the loss of precision on $g_{t\bar{t}H}$ measurement is negligible due to this background. An amendment in the analysis involves the $t\bar{t}$ background normalisation which is 5\% in our analysis. If the exact background normalisation ${\Delta\sigma_{\mathrm{eff}}^{\mathrm{BG}}}/{\sigma_{\mathrm{eff}}^{\mathrm{BG}}}$ is used, our results might change. Once the results from the group working on measuring the cross section of top pair production~\cite{ref6.4} are ready, they can be included in the analysis.


\begin{footnotesize}


\end{footnotesize}


\end{document}